\documentclass[preprintnumbers,amsmath,amssymb,nofootinbib]{revtex4}
\pdfoutput=1

\usepackage{graphicx}
\usepackage{amssymb}
\usepackage{booktabs}
\usepackage{color}
\usepackage{xspace}

\newcommand{\nn}{\nonumber}

\newcommand{\eq}[1]{\begin{equation} #1 \end{equation}}
\newcommand{\eqa}[1]{\begin{eqnarray} #1 \end{eqnarray}}

\newcommand{\re}{{\rm Re}}
\newcommand{\im}{{\rm Im}}

\def \re{\text{Re}}
\def \im{\text{Im}}

\def \hA {\hat{A}}
\def \Ab {\bar{A}}
\def \Pb {\bar{P}}
\def \hP {\hat{P}}

\begin{document}
\linespread{1.1}

\title{Exploiting the Symmetries of P and  S wave for $B \to K^* \mu^+\mu^-$}
%
\author{Lars Hofer and Joaquim Matias\bigskip}

\affiliation{Universitat Aut\`onoma de Barcelona, 08193 Bellaterra, Barcelona}

\begin{abstract}
After summarizing the current theoretical status of the  four-body decay $B\to K^*(\to K\pi)\mu^+\mu^-$, we apply the formalism of spin-symmetries to the full angular distribution, including the S-wave part involving a broad scalar resonance $K^*_0$. While we recover in the P-wave sector the known relation between the angular observables $P_i^{(\prime)}$, we find in the S-wave sector two new relations connecting the coefficients of the S-wave angular distribution and reducing the number of independent S-wave observables from six to four. Included in the experimental data analysis, these relations can help to reduce the background from S-wave pollution. We further point out the discriminative power of the maximum of the angular observable $P_2$ as a charm-loop insensitive probe of right-handed currents. Moreover, we show that in  absence of right-handed currents  the angular observables $P_4^\prime$ and $P_5^\prime$ fulfill the  relation $P_4^\prime= \beta P_5^\prime$ at the position where $P_2$ reaches its maximum.  
\end{abstract}
%

\maketitle                             
\section{Introduction}

Rare B decays constitute one of the cornerstones in the search for physics beyond the Standard Model (SM). Among them, the semileptonic mode $B \to K^*(\to K\pi) \mu^+\mu^-$ represents a particularly interesting channel as the measurement of the  4-body angular distribution provides a plethora of information which can be used to probe and discriminate different scenarios of New Physics (NP). In 2013, LHCb  presented results of the measurement of an optimized set $\{P_i^{(\prime)}\}$ of angular observables \cite{thefirst, Becirevic:2011bp,complete,implications,optimized} based on 1\,fb$^{-1}$  data. These observables are constructed in such a way that, to leading order in the strong coupling constant $\alpha_s$ and in the large-recoil expansion, non-perturbative form factors cancel in the region of low squared invariant mass $q^2$ of the dilepton pair, a unique and powerful feature in the hadronic environment.

Experimental data showed several interesting tensions with respect to SM expectations \cite{anomaly}: Most striking is the $4\,\sigma$ anomaly\footnote{In Ref.~\cite{explhcb} this discrepancy is quoted as a $3.7\,\sigma$ tension between the experimental result and the 68.3\% confidence level of the theoretical prediction, while we have quoted the tension between the experimental result and the theoretical central value. Note also that using the updated predictions \cite{impact} for all observables, including parametric and form factor errors,  factorizable power corrections together with an estimate of non-factorizable ones and charm-loop effects, the tensions with data, albeit slightly reduced, are still clearly present.} encountered in the observable $P_5^\prime$~\cite{implications} in the bin $[4.3,8.68]$\,GeV$^2$. The observable $P_2$~\cite{Becirevic:2011bp,complete} further displayed a $2.9\,\sigma$ deviation in the $q^2$-bin $[2,4.3]$\,GeV$^2$. The position of its zero ($q_0^2=4.9 \pm 0.9$ GeV$^2$), which is identical to the zero of the forward-backward asymmetry $A_{\rm FB}$, is in agreement with the SM prediction $q_0^2\simeq 4$ GeV$^2$ but allows for higher values. It is remarkable that all these deviations point to the same negative NP contribution $C_9^{\rm NP}$ to the Wilson coefficient of the semileptonic operator $\mathcal{O}_9$, possibly accompanied by a NP contribution $C_7^{\rm NP}$ to the Wilson coefficient of the magnetic operator $\mathcal{O}_7$. New Physics contributions to the Wilson coefficient $C_{10}$, and, in particular, to the coefficients  $C_{7,9,10}^\prime$ of the chirality-flipped operators are consistent with zero already at $1\,\sigma$. The full pattern, first pointed out in Ref.~\cite{anomaly} and obtained using all available experimental bins in $B\to K^*\mu^+\mu^-$ together with data on $B \to K^*\gamma$, $B \to X_s \gamma$, $B \to X_s \mu^+\mu^-$ and $B_s \to \mu^+\mu^-$, is given by the $1\,\sigma$ ranges
\begin{eqnarray} \label{pat}
C_9^{\rm NP}&\in&[-1.6,-0.9], \quad\quad\quad\;\, C_7^{\rm NP}\in[-0.05,-0.01], \quad\quad\quad 
C_{10}^{\rm NP}\in[-0.4,1.0], \nonumber\\ 
C_9^{\prime \rm NP}&\in&[-0.2,0.8], \quad\quad\quad\quad C_7^{\prime \rm NP}\in[-0.04,0.02],\quad\quad\quad\;\; C_{10}^{\prime \rm NP}\in[-0.4,0.4], 
\end{eqnarray}
where the mild preference for a positive $C_{10}^{\rm NP}$ is mainly driven by 
$B_s \to \mu^+\mu^-$ data.

The large negative NP contribution to $C_9$ was independently confirmed later on by other groups, using different observables $S_i$~\cite{buras,straub} (relying on the single large-recoil bin [1,6]\,GeV$^2$ and low recoil data), different statistical approaches \cite{bayesian} or form factor input from lattice \cite{meinelwingate}. Although it had been shown in Refs.~\cite{anomaly,proceedvip} that a large  $C_9^{\rm NP}+C_9^\prime<0$ was preferred in order to explain the $P_5^\prime$ anomaly, the possibility of a substantial positive $C_9^\prime$ enforcing $C_9^{\rm NP}+C_9^\prime\sim 0$ was discussed in Refs.~\cite{straub,hiller}, driven mainly by the 1 fb$^{-1}$ data  \cite{bkpp1fb} on the charged $B$ decay $B^+ \to K^+ \mu^+\mu^-$ in the region of low hadronic recoil. The situation has  become more coherent recently as the latest 3 fb$^{-1}$ data on $B^+ \to K^+ \mu^+\mu^-$ and $B^0 \to K^0 \mu^+\mu^-$ provided by LHCb \cite{bkpp3fb} is also in good agreement with the solution $C_9^{NP}+C_9^{\prime}<0$ \cite{anomaly}, both in the region of large  as well as  low hadronic recoil \cite{inprepa, wingate}. The three modes thus seem to point to a consistent overall picture of NP in agreement with the pattern
given by Eq.~(\ref{pat}). Moreover, under the assumption that NP affects only muons but not electrons, also the $2.6\,\sigma$ deviation measured by LHCb~\cite{lhcbrk} in the observable 
\begin{equation} R_K=\frac{{\text Br}(B^+\to K^+\mu^+\mu^-)}{{\text Br}(B^+\to K^+e^+e^-)}\end{equation} can be explained within the same scenario~\cite{ghosh,Hurth:2014vma,straub2}. In order to be able to draw solid conclusions and to see how this pattern evolves, it will  be crucial to know the 3\,fb$^{-1}$ data on the observables $P_i^{({\prime})}$ in $B \to K^*\mu^+\mu^-$. 

In parallel, the question has been raised if the observed discrepancies between data and SM predictions could be attributed to non-perturbative QCD effects~\cite{jaeger}, even though hadronic form factors enter optimized observables $P_i^{(\prime)}$ only at order $\alpha_s$ or through corrections breaking the large-recoil symmetries ({\it factorisable power corrections}). There exist two different approaches to account for factorisable power corrections: they can either be calculated (under certain modelling assumptions) within a non-perturbative framework like light-cone sum-rules (LCSR)~\cite{straub,StraubZwicky}, or they can be estimated exclusively on the basis of dimensional arguments and fundamental model-independent relations~\cite{jaeger,impact,jaeger2}. While the first method with full correlations among the form factors is suitable in order to extract the maximal information from a particular non-perturbative calculation, the second option in which correlations are included via large-recoil symmetry relations reduces the dependence on non-perturbative input to a minimum. The two approaches are thus complementary and, because the large-recoil symmetries are expected to be the dominant source of correlations, they should give similar results. Indeed, the resulting uncertainties obtained with the first method in Ref.~\cite{straub} and with the second method in Ref.~\cite{impact} are of the same order of 
magnitude (see also Ref.~\cite{proc}). Both these analyses find hadronic uncertainties from form factors to be under control\footnote{Refs.~\cite{jaeger,jaeger2}, on the other hand, quote much larger uncertainties. One of the reasons for that has been identified in Ref.~\cite{impact}: the decomposition of a form factor into a leading-order part and a $\mathcal{O}(\Lambda/m_b)$ power correction is not unique but (as in any fixed-order calculation) introduces a scheme dependence of observables at neglected higher orders in $\Lambda/m_b$. As the observables are effectively calculated at leading order ($\mathcal{O}(\Lambda/m_b)$ effects are not calculated but only estimated), they exhibit a scheme dependence at $\mathcal{O}(\Lambda/m_b)$ implying a $\sim 100\%$ scheme dependence of power corrections. In order to ensure predictivity of the method, it is hence crucial to exploit the freedom of choosing a scheme to minimize the impact of the unknown power corrections on the relevant observables (in the same way as in a fixed-order calculation in quantum field theory the renormalization scheme is chosen such that neglected higher orders do not spoil the perturbative expansion). It was demonstrated in Ref.~\cite{impact} that the sensitivity to factorizable 
power corrections of the key observables like $P_5^\prime$ is significantly reduced if a different scheme is chosen than the one employed in Refs.~\cite{jaeger,jaeger2}.}.

As a different explanation of the anomaly, the possibility of a large non-perturbative charm-loop contribution has been proposed \cite{zwicky}, requiring a huge correction with respect to theory predictions within the factorization approximation. The discussion in Ref.~\cite{zwicky} relies on two model-dependent assumptions: First that the resonance structure obtained from a fit to high-$q^2$ data on the scalar mode $B^+\to K^+\mu^+\mu^-$ can directly be transferred to the vector mode $B \to K^* \mu^+\mu^-$, and second that it can be extrapolated to low values of $q^2$. 
The only existing calculation \cite{khodjamirian} seems to be in contradiction with the low-$q^2$ scenarios of $B\to K^*\mu^+\mu^-$ obtained with this ansatz in Ref.~\cite{zwicky} as it finds a much smaller size for the charm loop and, moreover, the opposite sign for its contribution in $B^+\to K^+\mu^+\mu^-$ as compared to $B\to K^* \mu^+\mu^+$ (contrary to the assumption in Ref.~\cite{zwicky}).
Furthermore, if the 2.6$\sigma$ deviation in the observable $R_K$ persists, it poses a serious problem for the charm-loop or any other low-energy QCD explanation which cannot generate effects violating lepton-flavor universality. Also the observable $P_2$ in $B\to K^* \mu^+\mu^+$ can be instrumental in testing the charm-loop hypothesis proposed in Ref.~\cite{zwicky} (see also \cite{proc}).

While the polluting effects from non-perturbative QCD have been studied in detail in the literature, less attention has been paid to the so-called S-wave pollution, generated by the background decay $B \to K^*_0(\to K\pi) \mu^+\mu^-$ where $K^*_0$ is a broad scalar resonance. In Ref.~\cite{becirevicswave} a detailed and complete calculation of the S-wave background was performed and it was concluded that any observable will unavoidably suffer from its pollution. While this conclusion is correct in the case of uniangular distributions, it does not apply to full or folded distributions where the P- and the S-wave parts can be separated according to their different angular dependence. As shown in Ref.~\cite{swave}, S-wave pollution can  be avoided for the $P_i^{(\prime)}$ observables if folded distributions are used instead of uniangular ones. A discussion of the experimental implications of the S-wave contribution was presented in Ref.~\cite{ulrikswave} (see also Ref.~\cite{julich}). 

Experimental analyses of $B\to K^*\mu^+\mu^-$ rely on theoretical information regarding the S-wave background. To this end, a set of model-independent bounds on the coefficients of the S-wave part of the angular distribution was presented in Ref.~\cite{optimized}, derived from application of the Cauchy-Schwarz inequality. On the other hand, it was shown in Refs.~\cite{ulrik1,complete} that the coefficients of the P-wave part are not independent parameters but that they are correlated through the spin-symmetry of the angular distribution. In this work we transfer this idea to the S-wave sector. We derive two relations which effectively reduce the number of free coefficients of the S-wave distribution from six to four. It is expected that the inclusion of these relations into the data analysis can help to further improve the background estimation. We illustrate the effect of the correlations for the ratio of the S-wave observables $A_S^4$ and $A_S^5$ and study implications at the position $q^2=q_1^2$ of the maximum  of the observable $P_2$. Moreover, we point out a relation between $P_4^\prime$ and $P_5^\prime$ at $q^2=q_1^2$ and suggest to use the  maximum of $P_2$ as a golden observable to probe right-handed currents (for an explicit model generating right-handed currents see e.g. Ref.~\cite{lunghi}).   

The outline of this paper is the following: In Sec.~\ref{sec:SpinSym} we discuss the spin-symmetry of the differential $B\to K^*\mu^+\mu^-$ decay rate and determine the number of independent observables in the P- and in the S-wave sector. In Sec.~\ref{sec:SymRel} we derive the resulting symmetry relations. Their phenomenological consequences are discussed in Sec.~\ref{sec:Pheno}. First, we study the discriminating power of the maximum of $P_2$ as a test for right-handed currents, then we determine a relation between $P_4^\prime$ and $P_5^\prime$ at the position of the maximum of $P_2$, and finally we investigate constraints on the S-wave observables $A_S^{(i)}$ and derive simple relations among them at the position of the maximum and the zero of $P_2$. Sec.~\ref{sec:Conclu}  contains our conclusions. In Appendix~A we present an explicit example of how to use the freedom introduced by the symmetries to fix a possible convention for the amplitudes, while Appendix~B contains details of the derivation of the bound on $A_S$.

\section{Spin symmetry of the differential decay rate}
\label{sec:SpinSym}

The differential decay rate of the full four-body decay $B\to K\pi \ell^+\ell^-$ receives contributions from the P-wave decay $B\to K^*(\to K\pi) \ell^+\ell^-$ as well as from the S-wave decay $B\to K_0^*(\to K\pi) \ell^+\ell^-$ with $K_0^*$ being a broad scalar resonance. It can thus be decomposed into a P-wave and an S-wave part,
\begin{equation}
\frac{d^5\Gamma}{dq^2\,dm_{K\pi}^2\,d\!\cos\theta_K\,d\!\cos\theta_\ell\,d\phi}\,=\, W_P\,+\,W_S,
\end{equation}
with $W_P$ containing the pure P-wave contribution and $W_S$ containing the contributions from pure S-wave exchange as well as from S-P interference. Here, $q^2$ denotes the square of the invariant mass of the lepton pair and $m_{K\pi}$ the invariant mass of the $K\pi$ system. Further, $\theta_\ell$, $\theta_K$ are the angles describing the relative directions of flight of the final-state particles, while $\phi$ is the angle between the dilepton and the dimeson plane (see Ref.~\cite{ulrik1} for  exact definitions). Angular momentum conservation dictates the dependence of $W_P$ and $W_S$ on $\theta_\ell$, $\theta_K$, $\phi$ to be  
\begin{eqnarray}
W_P&=&\frac9{32\pi} \left[
J_{1s} \sin^2\theta_K + J_{1c} \cos^2\theta_K + (J_{2s} \sin^2\theta_K + J_{2c} \cos^2\theta_K) \cos 2\theta_l\right.\nn\\[1.5mm]
&&\hspace{1cm}+ J_3 \sin^2\theta_K \sin^2\theta_l \cos 2\phi + J_4 \sin 2\theta_K \sin 2\theta_l \cos\phi  + J_5 \sin 2\theta_K \sin\theta_l \cos\phi \nn\\[1.5mm]
&&\hspace{1cm}+ (J_{6s} \sin^2\theta_K +  {J_{6c} \cos^2\theta_K})  \cos\theta_l    
+ J_7 \sin 2\theta_K \sin\theta_l \sin\phi  + J_8 \sin 2\theta_K \sin 2\theta_l \sin\phi \nn\\[1.5mm]
&&\hspace{1cm}\left.+ J_9 \sin^2\theta_K \sin^2\theta_l \sin 2\phi \right]
\end{eqnarray}
and
\begin{eqnarray}
W_S &=&\frac{1}{4\pi} \left[ {\tilde J}_{1a}^c+{\tilde J}_{1b}^c \cos \theta_K+ ({\tilde J}_{2a}^c+{\tilde J}_{2b}^c \cos\theta_K ) \cos 2\theta_\ell + {\tilde J}_{4} \sin \theta_K \sin 2 \theta_\ell \cos \phi \right.\nn\\ &&\left.\hspace{0.7cm}+
{\tilde J}_{5} \sin \theta_K \sin \theta_\ell \cos \phi + {\tilde J}_{7} \sin \theta_K \sin \theta_\ell \sin \phi+
{\tilde J}_{8} \sin \theta_K \sin 2\theta_\ell \sin \phi\right].
\end{eqnarray}
The coefficients $J_i$ and $\tilde J_i$ are functions of  $q^2$ and $m_{K\pi}^2$.

If not explicitly stated otherwise, we will neglect lepton masses in the following. Then, the decays $B\to K^*\ell^+\ell^-$ and $B\to K_0^*\ell^+\ell^-$ are described by six complex amplitudes $A_{\|,\bot,0}^{L,R}$ and two complex amplitudes $A_0^{\prime L,R}$, respectively, where the upper index $L,R$ refers to the chirality of the outgoing lepton current, while in the case of the P-wave the lower index $\|,\bot,0$ indicates the helicity of the $K^*$-meson\footnote{In the case of non-vanishing lepton masses and of scalar operators coupling to the lepton pair, two additional amplitudes $A_t$ and $A_S$ have to be included.}. These amplitudes are multiplied by a Breit-Wigner propagator
$BW_i(m_{K\pi}^2)$ with $i=K^*,K^*_0$
describing the propagation of the $K^*$ and $K_0^*$ meson, respectively. For the exact form of the Breit-Wigner functions $BW_i(m_{K\pi}^2)$ we refer to Ref.~\cite{becirevicswave}.

Since the final state is summed over the spins of the leptons, the obervables $J_i$ and $\tilde J_i$ are exclusively described in terms of spin-summed squared amplitudes of the form $A_i^{L*}A_j^{L}\pm A_i^{R*}A_j^{R}$ \footnote{Interferences of different $K^*$ and $K_0^*$ helicities $i\not=j$ contribute, as these particles only appear as unobserved intermediate states.}. This pattern suggests to combine left- and right-handed amplitudes to two-component complex vectors:
\eq{\label{eq:nvecs}
n_\|=\binom{A_\|^L BW_P}{A_\|^{R*} BW_P^*}\ ,\quad
n_\bot=\binom{A_\bot^L BW_P}{-A_\bot^{R*} BW_P^*}\ ,\quad
n_0=\binom{A_0^L BW_P}{A_0^{R*} BW_P^*}\ ,\quad
n_S=\binom{A_0^{\prime L} BW_S}{A_0^{\prime R*} BW_S^*}\ . 
}
Using this notation, the observables $J_i$ and $\tilde J_i$ can be expressed in terms of scalar products $n_i^\dagger n_j$ of these vectors. Neglecting lepton masses and presence of scalars we find
\begin{eqnarray}
J_{1s} & = & \frac34\left( |n_\bot|^2+ |n_\||^2\right),\hspace{0.9cm}
J_{1c}  \,=\,|n_0|^2 \,, \hspace{3cm}
J_{2s}  \,=\, \frac14 \left(|n_\bot|^2+ |n_\||^2 \right) \,, \nn\\[1mm]
J_{2c} & = & -|n_0|^2 \,,\hspace{2.85cm}
J_3  \,=\, \frac12 \left( |n_\bot|^2 - |n_\||^2 \right)\,, \hspace{0.9cm}
J_4  \,=\, \frac1{\sqrt{2}} \re (n_0^\dagger\,n_\|)\,,\nn\\[1mm]
J_5 & = &  \sqrt{2}\, \re (n_0^\dagger\,n_\bot)\,, \hspace{1.45cm}
J_{6s}  \,=\, 2\, \re (n_\bot^\dagger\,n_\|)\,, \hspace{1.85cm}
J_{7} \,=\,  -\sqrt{2}\, \im (n_0^\dagger\,n_\|)\,,\nn \\[1mm]
J_8 & \,=\, & -\frac1{\sqrt{2}} \im (n_0^\dagger\,n_\bot) \,,\hspace{1.25cm}
J_9 \,=\, -\im (n_\bot^\dagger\,n_\|)\,,\hspace{1.65cm}
J_{6c} \,=\, 0\,,
\end{eqnarray} 
and
\eqa{
{\tilde J_{1a}^c}=-{\tilde J_{2a}^c}&=&\frac{3}{8} (|A_0'^L|^2 +| A_0'^R|^2)|BW_S|^2= \frac{3}{8}|n_S|^2, \nn \\
{\tilde J_{1b}^c}=-{\tilde J_{2b}^c}&=& \frac{3}{4}\sqrt{3}   {\rm Re} \left[(
A_0'^L A_0^{L*} + A_0'^R A_0^{R*})BW_S BW_P^*
  \right]= \frac{3}{4}\sqrt{3} {\rm Re} (n_S^\dagger\, n_0), \nn \\
{\tilde J_4}&=& \frac{3}{4}\sqrt{\frac{3}{2}}  {\rm Re} \left[(
A_0'^L A_\|^{L*} + A_0'^R A_\|^{R*})BW_S BW_P^*
\right]=  \frac{3}{4}\sqrt{\frac{3}{2}} {\rm Re} (n_S^\dagger\, n_\|),\nn \\
{\tilde J_5}&=&
\frac{3}{2}\sqrt{\frac{3}{2}} {\rm Re} \left[(
A_0'^L A_\perp^{L*} - A_0'^R A_\perp^{R*})BW_S BW_P^*
\right]=\frac{3}{2}\sqrt{\frac{3}{2}} {\rm Re} (n_S^\dagger\, n_\perp), \nn \\
{\tilde J_7}&=& \frac{3}{2}\sqrt{\frac{3}{2}}   {\rm Im} \left[(
A_0'^L A_\|^{L*} - A_0'^R A_\|^{R*})BW_S BW_P^*
\right]=\frac{3}{2}\sqrt{\frac{3}{2}} {\rm Im} (n_\|^\dagger\, n_S), \nn \\
{\tilde J_8}&=&\frac{3}{4}\sqrt{\frac{3}{2}}  {\rm Im} \left[(
A_0'^L A_\perp^{L*} + A_0'^R A_\perp^{R*})BW_S BW_P^*
  \right]=\frac{3}{4}\sqrt{\frac{3}{2}} {\rm Im} (n_\perp^\dagger n_S).}

The fact that the $J_i$ and $\tilde J_i$ observables involve a sum over the spins of the leptons implies that they are not sensitive to the full information contained in the helicity amplitudes $A_{\|,\bot,0}^{L,R}$, $A_0^{\prime L,R}$. This can be easily seen from the notation in terms of the vectors $n_i$. As the $J_i$ and $\tilde J_i$ observables are scalar products of the $n_i$, they are invariant under a $U(2)$ rotation of these vectors. It is thus impossible to fully reconstruct the amplitudes from the $J_i,\tilde J_i$ observables alone. If one wishes to extract the $A_{\|,\bot,0}^{L,R}$, $A_0^{\prime L,R}$ from experiment, it is mandatory to fix a convention which resolves the ambiguity related to the $U(2)$ symmetry. A possible choice is presented in Appendix A.

The number of independent observables that can be constructed from $n_A$ complex amplitudes is given by $2n_A$. In presence of a symmetry $S$ with $n_{\rm gen}$ generators, there exist 
\eq{n_{\rm Obs}\,=\,2n_A-n_{\rm gen}} 
independent observables which respect the symmetry $S$. The $U(2)$ spin symmetry of the $J_i$ and $\tilde J_i$ observables with $n_{\rm gen}=4$ generators thus leads to the following consequences:
\begin{itemize}
  \item In the P-wave sector there are $n^P_{\rm Obs}=2\cdot 6-4=8$ independent observables. This observation implies the existence of a relation between the 9 non-trivially different P-wave coefficients $J_i$. The corresponding relation has been derived in Ref.~\cite{ulrik1} and its phenomenological consequences have been discussed in Ref.~\cite{nicola}. 
  \item In the S-wave sector there are $n^S_{\rm Obs}=2\cdot 8-4-n^P_{\rm Obs}=4$ additional observables. This observation implies the existence of two additional relations among the 6 S-wave coefficients $\tilde J_i$ and the P-wave coefficients $J_i$. These relations will be derived in the following section.
\end{itemize}
This parameter counting implies that a basis in the P-wave sector consists of 8 independent observables, like the basis $\{\Gamma^\prime,A_{\rm FB}\, {\rm or}\, F_L,P_1,P_2,P_3,P_4^\prime,P_5^\prime,P_6^\prime\}$ proposed in Ref.~\cite{optimized}. In particular, the observables of this basis are not related among each other through a symmetry, but they are connected to the observable $P_8^\prime$. In the S-wave sector, a basis consists of 4 independent observables. This means that from the complete set of S-wave observables $\{F_S,A_S,A_S^4,A_S^5,A_S^7,A_S^8\}$ (see Sec.~\ref{sec:SymRel} for their definition) a subset of four has to be chosen as basis, while the remaining two are obtained from symmetry relations.

\section{P-wave and S-wave symmetry relations}
\label{sec:SymRel}
The observables $J_i$ and $\tilde J_i$ can be expressed in terms of scalar products $n_i^\dagger n_j$. Since $n_\|$ and $n_\bot$ span the space of complex 2-component vectors, the other two vectors can be expressed as linear combinations of the former:
\eq{\label{eq:decomp}
n_i=a_in_\|+b_in_\bot,\quad i=0,S. 
}
Contracting with $n_\|$ and $n_\bot$ we get a system of linear equations
\begin{eqnarray}
  n_\|^\dagger n_i&=&a_i|n_\||^2+b_i (n_\|^{\dagger} n_\bot), \nonumber\\
  n_\bot^\dagger n_i&=&a_i (n_\bot^{\dagger} n_\|) + b_i |n_\bot|^2,
\end{eqnarray}
which can easily be solved for $a_i,b_i$:
\begin{eqnarray}
  a_i=\frac{|n_\bot|^2(n_\|^\dagger n_i)-(n_\|^\dagger n_\bot)(n_\bot^\dagger n_i)}{|n_\||^2|n_\bot|^2-|n_\bot^\dagger n_\||^2},\qquad
  b_i=\frac{|n_\||^2(n_\bot^\dagger n_i)-(n_\bot^\dagger n_\|)(n_\|^\dagger n_i)}{|n_\||^2|n_\bot|^2-|n_\bot^\dagger n_\||^2}.
\end{eqnarray}
Using the decomposition (\ref{eq:decomp}) of $n_0,n_S$ in terms of $n_\|,n_\bot$ to calculate the scalar products $|n_0|^2,|n_S|^2,n_0^\dagger n_S$, one finds
\begin{eqnarray}
  |n_i|^2&=&a_i(n_i^\dagger n_\|)+b_i(n_i^\dagger n_\bot),\quad (i=0,S) \nonumber\\
  n_0^\dagger n_S&=&a_S(n_0^\dagger n_\|)+b_S(n_0^\dagger n_\bot).
\end{eqnarray}
Reexpressed in terms of the coefficients $J_i$, $\tilde J_i$ of the angular distribution, this gives the three symmetry relations\footnote{The same results are obtained if instead of $\{n_\|,n_\bot\}$ a different subset $\{n_i,n_j\}$ is chosen as basis and the derivation is adjusted accordingly. In particular, the stated results are valid also for values of $q^2$ for which $n_\|$ and $n_\bot$ become aligned.}:
\begin{eqnarray}
J_{2c}\left[16 J_{2s}^{2} -
\left(4 J_3^2+ J_{6s}^{2} + 4 J_9^2 \right)\right] &= & \ 4 \left[J_{6s} (J_4 J_5 +
 J_7 J_8) + J_9 (J_5 J_7 - 4 J_4 J_8)\right]\nonumber\\[2ex]
&&
\ -2\left[(2 J_{2s}+ J_3) \left(4 J_4^2+J_7^2\right)
+ ( 2 J_{2s} - J_3)
\left(J_5^2+4 J_8^2 \right)\right],\label{eq:Psym} \\[4ex] 
  -\frac{9}{2}{\tilde J^c_{1a}}\left[16 J_{2s}^{2} -
\left(4 J_3^2+ J_{6s}^{2} + 4 J_9^2 \right)\right]&=&\ 4 \left[J_{6s} ({\tilde J}_4 {\tilde J}_5 +
 {\tilde J}_7 {\tilde J}_8) + J_9 ({\tilde J}_5 {\tilde J}_7 - 4 {\tilde J}_4 {\tilde J}_8)\right]\nonumber\\[2ex]
&&
\ -2\left[(2 J_{2s}+ J_3) \left(4 {\tilde J}_4^2+{\tilde J}_7^2\right)
+ ( 2 J_{2s} - J_3)
\left({\tilde J}_5^2+4 {\tilde J}_8^2 \right)\right],\label{eq:Ssym1} \\[4ex]
  2{\tilde J}^c_{1b}\left[16J_{2s}^2-(4J_3^2+J_{6s}^2+4J_9^2)\right]&=&-4\left[J_{6s}(J_4{\tilde J}_5+J_5{\tilde J}_4+J_7{\tilde J}_8+J_8{\tilde J}_7)+J_9(J_5{\tilde J}_7+J_7{\tilde J}_5-4J_4{\tilde J}_8-4J_8{\tilde J}_4)\right]
  \nonumber\\[2ex]
    &&+4\left[(2J_{2s}+J_3)(4J_4{\tilde J}_4+J_7{\tilde J}_7)+(2J_{2s}-J_3)(J_5{\tilde J}_5+4J_8{\tilde J}_8)\right].\label{eq:Ssym2}
\end{eqnarray}
Eq.~(\ref{eq:Psym}) had already been derived in Ref.~\cite{ulrik1}, determining explicitly the amplitudes in terms of the $J_i$ coefficients after fixing a ``gauge convention'' (see Appendix A for a possible gauge condition). Here, it has been obtained in a ``gauge-independent'' way. As a cross-check, we have also rederived Eqs.~(\ref{eq:Ssym1}),(\ref{eq:Ssym2}) following the alternative procedure of Ref.~\cite{ulrik1}. Of the two relations involving S-wave parameters, eq.~(\ref{eq:Ssym1}) and eq.~(\ref{eq:Ssym2}), the first one is quadratic in the ${\tilde J}_i$ while the second one is linear. It is interesting to note that relation (\ref{eq:Ssym1}) for the S-wave coefficients ${\tilde J}_{4,5,7,8}$ has the same structure as the well-known relation (\ref{eq:Psym}) for the P-wave coefficients $J_{4,5,7,8}$, and further that the combination of the three equations for $J_{2c} -\frac{9}{2} {\tilde J}_{1a}^c \mp 2 {\tilde J}_{1b}^c$ has exactly the same structure as Eq.(\ref{eq:Psym}) substituting $J_i \to J_i\pm{\tilde J}_i$ for $i=4,5,7,8$. 

The S-wave observables are defined as
\begin{eqnarray}
A_S=\frac{8}{3} \frac{\tilde J_{1b}^c + \bar{\tilde J}_{1b}^c}{\Gamma_{\rm full}^\prime +\overline{\Gamma}_{\rm full}^\prime}, \hspace{2cm}&& 
A_S^{\rm CP}=\frac{8}{3} \frac{\tilde J_{1b}^c - \bar{\tilde J}_{1b}^c}{\Gamma_{\rm full}^\prime +\overline{\Gamma}_{\rm full}^\prime},\nn\\ 
A_S^i=\frac{4}{3} \frac{\tilde J_i + \bar{\tilde{J}}_i}{\Gamma_{\rm full}^\prime+\overline{\Gamma}_{\rm full}^\prime},\hspace{2cm} &&
A_S^{i\rm{\,CP}}=\frac{4}{3} \frac{\tilde J_i - \bar{\tilde{J}}_i}{\Gamma_{\rm full}^\prime+\overline{\Gamma}_{\rm full}^\prime},
\end{eqnarray}
where $\bar J_i$, $\bar{\tilde J}_{i}$ and $\overline{\Gamma}_{\rm full}^\prime$ denote the corresponding angular coefficients and differential decay width for the CP-conjugated decays $\bar{B}\to \bar{K}^*\mu^+\mu^-$ and $\bar{B}\to \bar{K}^*_{0}\mu^+\mu^-$. The total differential decay width $\Gamma_{\rm full}^\prime$ is given by
\eq{\Gamma^{\prime}_{\rm full}\,=\,\Gamma^{\prime}_{K^*}+\Gamma^{\prime}_{K^*_0},}
where in the limit of massless leptons
\eq{\Gamma^{\prime}_{K^*}=4J_{2s}-J_{2c},\hspace{0.5cm}\Gamma^{\prime}_{K^*_0}=\frac{8}{3}{\tilde J}^c_{1a}.}
Expressing Eqs.~(\ref{eq:Psym})-(\ref{eq:Ssym2}) in terms of the S-wave observables $A_S^{(i)}$ and
\begin{equation}F_S=\frac{\Gamma^{\prime}_{K^*_0}+\bar\Gamma^{\prime}_{K^*_0}}{\Gamma^{\prime}_{\rm full} + \bar \Gamma^{\prime}_{\rm full}}\end{equation}
and the P-wave observables $P_i^{(\prime)}$ and $F_T$ (as defined e.g. in Ref.~\cite{optimized}) we obtain
\begin{eqnarray}
   {k}_L\left[k_T^2-P_1^2-4P_2^2-4P_3^2\right]&=&-4P_2\left[P_4^\prime P_5^\prime+P_6^\prime P_8^\prime\right]
       -4P_3\left[P_5^\prime P_6^\prime-P_4^\prime P_8^\prime\right]\nonumber\\
       &&+(k_T+P_1)\left[(P_4^\prime)^2+(P_6^\prime)^2\right]+(k_T-P_1)
       \left[(P_5^\prime)^2+(P_8^\prime)^2\right],\label{eq:PRel}\\   
   {k}_S F_T F_S (1-F_S)\left[k_T^2-P_1^2-4P_2^2-4P_3^2\right]&=&-\frac{8}{3}P_2\left[A_S^4A_S^5+A_S^7A_S^8\right]
       +\frac{4}{3}P_3\left[A_S^5A_S^7-4A_S^4A_S^8\right]\nonumber\\
       &&+\frac{1}{3}(k_T+P_1)\left[4(A_S^4)^2+(A_S^7)^2\right]+\frac{1}{3}(k_T-P_1)
       \left[(A_S^5)^2+4(A_S^8)^2\right],\;\;\;\;\label{eq:AsRel1}\\
   A_S\sqrt{\frac{F_T}{1-F_T}}\left[k_T^2-P_1^2-4P_2^2-4P_3^2\right]&=&-4P_2
       \left[P_4^\prime A_S^5+2P_5^\prime A_S^4-2P_6^\prime A_S^8-P_8^\prime A_S^7\right]\nonumber\\
       &&+4P_3\left[P_5^\prime A_S^7-P_6^\prime A_S^5-2P_4^\prime A_S^8+2P_8^\prime A_S^4\right]\nonumber\\
       &&+2(k_T+P_1)\left[2P_4^\prime A_S^4-P_6^\prime A_S^7\right]
         +2(k_T-P_1)\left[P_5^\prime A_S^5-2P_8^\prime A_S^8\right],\label{eq:AsRel2}
\end{eqnarray}
with
\begin{equation}
  k_L=k_T=k_S=1.
\end{equation}
These relations are valid up to terms which are quadratic in the CP-violating parameters $A_S^{(i){\rm CP}}$, $F^{\rm CP}_S$, $P_i^{(\prime)\,{\rm CP}}$ and $F_T^{{\rm CP}}$. Exact versions of the equations can be obtained by the replacements
\begin{eqnarray}
   P_i^{(\prime)}&\to&\Pb_i^{(\prime)}=P_i^{(\prime)}+P_i^{(\prime)\,{\rm CP}},\hspace{2cm}
   A_S^{(i)\,{\rm CP}}\to\Ab^{(i)}_S\,=\,A_S^{(i)}+A_S^{(i)\,{\rm CP}},\nonumber\\
   k_i&\to&\bar{k}_i=1+F^{\rm CP}_i/F_i\hspace{1cm}(i=L,T,S),\label{eq:ExactRep1}
\end{eqnarray}
or
\begin{eqnarray}
   P_i^{(\prime)}&\to&\hP_i^{(\prime)}=P_i^{(\prime)}-P_i^{(\prime)\,{\rm CP}},\hspace{2cm}
   A_S^{(i)\,{\rm CP}}\to\hA^{(i)}_S\,=\,A_S^{(i)}-A_S^{(i)\,{\rm CP}},\nonumber\\
   k_i&\to&\hat{k}_i=1-F^{\rm CP}_i/F_i\hspace{1cm}(i=L,T,S).\label{eq:ExactRep2}
\end{eqnarray} 
In the form the equations are displayed, lepton masses are neglected. For the P-wave observables, the full lepton-mass dependence can easily be restored by the replacements $P_2\to\beta P_2$, $P_5^\prime\to\beta P_5^\prime$ and $P_6^\prime\to\beta P_6^\prime$ with $\beta=\sqrt{1-4m_\ell^2/q^2}$. In the following we will typically suppress factors of $\beta\approx 1$ and only restore them in final results. For the S-wave observables, given their poor experimental precision, we will neglect any terms suppressed by small lepton masses throughout the paper.   

Note that Eq.~(\ref{eq:PRel}) is equivalent to Eq.~(4) of Ref.~\cite{nicola}, while Eqs.~(\ref{eq:AsRel1}),(\ref{eq:AsRel2}) involving the S-wave parameters constitute the main result of the present work. The information contained in the two S-wave relations is twofold. On the one hand, they can be used to obtain independent bounds on the five S-wave observables $A_S$,$A_S^4$,$A_S^5$, $A_S^7$,$A_S^8$. As we will show, the resulting bounds are equivalent to the ones derived from the Cauchy-Schwarz inequality in Ref.~\cite{optimized}. On the other hand, the equations relate the six S-wave observables $A_S$,$A_S^4$,$A_S^5$, $A_S^7$,$A_S^8$ and $F_S$ to each other, reducing the number of independent observables effectively from six to four. These correlations should thus be implemented in the experimental data analysis in order to improve the background estimation.

\section{Phenomenological implications}
\label{sec:Pheno}
\boldmath
\subsection{Connecting $P_1$ and $P_2$: 
The maximum of $P_2$ as a test for the presence of RH currents}
\unboldmath\label{subsec:MaxP2}
Before discussing the phenomenological consequences of Eqs.~(\ref{eq:PRel})-(\ref{eq:AsRel2}), let us first have a closer look at the observable
\begin{equation}
   x=k_T^2-P_1^2-4\beta^2P_2^2-4P_3^2
   \label{eq:obsx}
\end{equation}
appearing on the left-hand side of these equations. In eq.~(\ref{eq:obsx}) we have reinstalled the dependence on the lepton mass by means of the parameter $\beta=\sqrt{1-4m_\ell^2/q^2}$. Expressing $k_T$,$P_{1,2,3}$ in terms of $n_\|$ and $n_\bot$ (and the respective vectors $\bar n_\|$ and $\bar n_\bot$ parametrising the CP conjugated amplitude), it can easily be shown that $x\ge 0$ up to terms quadratic in the CP-violating observables\footnote{The observables $\bar{x}$ and $\hat{x}$ constructed from eq.~(\ref{eq:obsx}) via the replacements (\ref{eq:ExactRep1}) and (\ref{eq:ExactRep2}), respectively, fulfill $\bar{x}\ge 0$ and $\hat{x}\ge 0$ exactly.} $F_T^{\rm CP}$,$P_1^{\rm CP}$,$P_2^{\rm CP}$,$P_3^{\rm CP}$. From this observation the upper bounds $|P_1|\le 1$, $|P_2|\le 1/(2\beta)$ and $|P_3|\le 1/2$ can be read off immediately. On the other hand, it can be concluded that, if one of the three observables $P_{1,2,3}$ saturates its bound at a point $q^2=q_1^2$, the other two observables have to vanish at this point. The experimental result $\langle P_2\rangle_{[2,4.3]}=0.50^{+0.00}_{-0.07}$ indeed suggests a quasi-saturation\footnote{Note that in practice a complete saturation cannot be accomplished due to the finite bin-size.} of the bound for the observable $P_2$ in the bin $[2,4.3]$ GeV$^2$. Depending on how this result evolves with the new data, the correlation with $P_1$ via 
the positivity condition $x\ge 0$ could be useful to constrain the less precisely measured observable $P_1$ in the respective bin.

In order to study the information encoded in the maximum of $P_2$ and the relation with the observable $P_1$ in more detail\footnote{We will assume real Wilson coefficients for all this discussion.}, let us have a look at the expressions of these observables in terms of the vectors $n_\perp$ and $n_\|$:
\begin{equation} \label{eq:P12}P_1=\frac{|n_\perp|^2-|n_\||^2}{|n_\perp |^2+ |n_\||^2},\hspace{2cm} 
P_2 = \frac{1}{2\beta}\left[1-\frac{(n_\perp-n_\|)^\dagger(n_\perp-n_\|)}{|n_\perp |^2+ |n_\||^2}\right].\end{equation}
Obviously, $P_2$ reaches the extreme value $1/(2\beta)$ at the position $q_1^2$ of its maximum if and only if $n_\perp(q_1^2)=n_\|(q_1^2)$, i.e. if $A_\perp^L(q_1^2)=A_\|^L(q_1^2)$ and $A_\perp^R(q_1^2)=-A_\|^R(q_1^2)$. At leading order, the second of these two conditions is automatically fulfilled in the absence of right-handed currents $C_7^\prime=C_9^\prime=C_{10}^\prime=0$, while the first condition is fulfilled in this case (and neglecting the small ${\rm Im}C_9^{\rm eff}$ entering $P_2$ quadratically) for  
\begin{equation} 
\label{eq:MaxPos}
q_1^2= \frac{2 m_b M_B C_7^{\rm eff}}{C_{10} - {\rm Re}\,C_9^{\rm eff}(q_1^2)}.
\end{equation}
From this observation we conclude that any CP-conserving new-physics contribution added to the Wilson-coefficients $C_{7,9,10}$ will shift the position $q_1^2$ of the maximum of $P_2$, while maintaining its height at $P_2^{\rm max}\sim 1/(2\beta)$. Compared to the SM-prediction $q_1^2\approx 2\,{\rm GeV}^2$, the experimental result $\langle P_2\rangle_{[2,4.3]}=0.50^{+0.00}_{-0.07}$ prefers a larger value for $q_1^2$, more to the center of the bin $[2,4.3]$ GeV$^2$. This pull to larger $q^2$-values for the position of the maximum of $P_2$ is consistent with the pull to larger $q^2$-values of its zero mentioned in the introduction. From Eq.~(\ref{eq:MaxPos}) we see that a larger $q_1^2$ can be obtained by a negative NP contribution to $C_9$, as required by the $P_5^\prime$ anomaly, and/or by a positive contribution to $C_{10}$. Notice further that, while it was claimed in Ref.~\cite{zwicky} that charm-loop effects might affect the position of the zero of $P_2$, their impact on the position of the maximum is basically negligible for all scenarios studied in Ref.~\cite{zwicky}. In general, the maximum of $P_2$
probes the Wilson coefficient $C_9^{\rm eff}$ in a different region in $q^2$ than the $P_5^\prime$ anomaly or
the zero of $P_2$. While a potential NP contribution to $C_9^{\rm eff}$ is $q^2$-independent and thus induces
exactly related effects in the three observables, a charm-loop contribution enters $C_9^{\rm eff}$ as a
non-trivial function of $q^2$ which is expected to decrease with increasing distance to the $c\bar{c}$ resonance
region. A measurement of the maximum of $P_2$ can thus help to discriminate between NP at high energies and non-perturbative charm effects, and the upcoming data with smaller-sized bins will help to determine it more precisely.

In contrast to $C_{7,9,10}$, a new right-handed contribution to one of the Wilson coefficients $C_{7,9,10}^\prime$ will not only shift the position $q_1^2$ of the maximum of $P_2$ but will also lower its value $P_2^{\rm max}$, pushing it  below $1/(2\beta)$ \cite{Becirevic:2011bp}. At leading order, this can be seen from the fact that in the presence of right-handed currents the identity $A_\perp^R=-A_\|^R$ does not hold anymore for all $q^2$ so that the two conditions $A_\perp^L(q_1^2)=A_\|^L(q_1^2)$ and $A_\perp^R(q_1^2)=-A_\|^R(q_1^2)$ required for $P_2(q_1^2)=1/(2\beta)$ cannot be fulfilled at the same point $q_1^2$. In general, right-handed currents will cause $|n_\perp(q_1^2)|\not=|n_\|(q_1^2)|$ and thus induce a substantial non-vanishing $P_1(q_1^2)$, preventing $P_2(q_2^2)$ from reaching its absolute maximum $1/(2\beta)$.
\begin{figure}
\includegraphics[width=8cm,height=5cm]{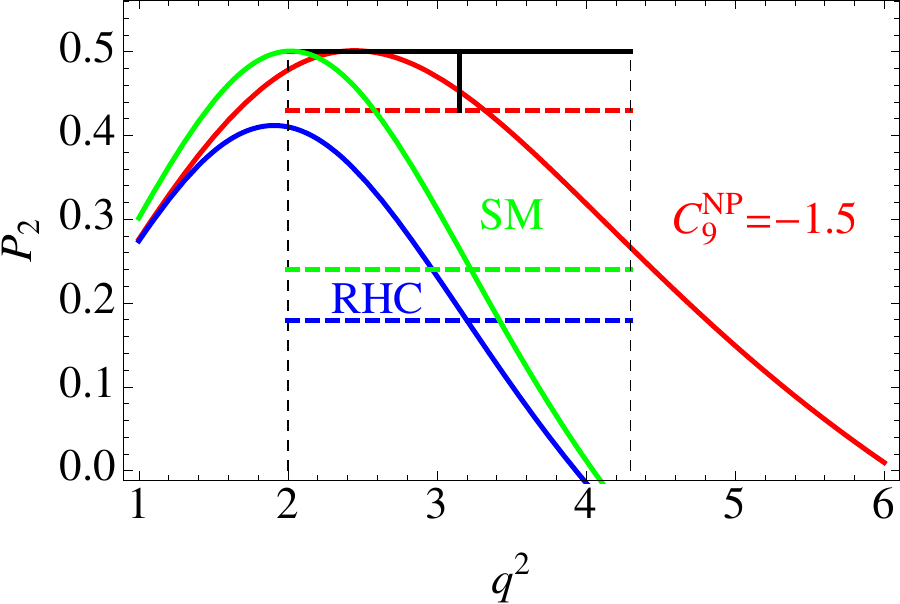}\hspace{1.5cm}
\includegraphics[width=8cm,height=5cm]{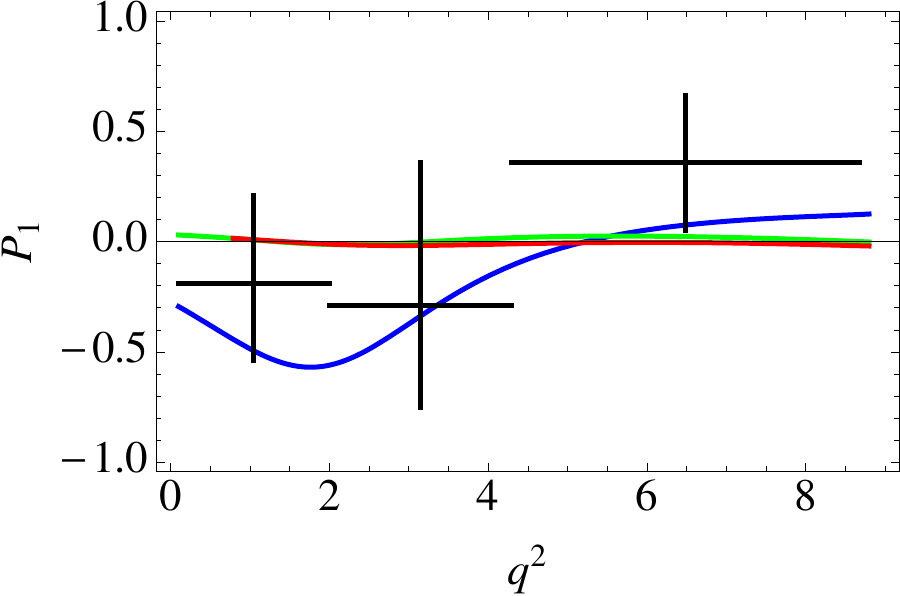}
\caption{\label{fig:P2max}Left: Comparison of the $P_2$-curves (central values) in the SM (green) and in two scenarios of New Physics. The scenario NP (red) corresponds to $C_9^{\rm NP}=-1.5$, the scenario RHC (blue) corresponds to $C_9^\prime=1$, $C_{10}^\prime=0.4$, $C_7^\prime=0.06$. Dashed lines represent the central value for the integrated bin $[2,4.3]$ GeV$^2$ of the respective curve, while the black cross indicates the measured value in this bin. Right: The analogous curves for $P_1$ with the black crosses representing the measured values in the respective bins.}
\end{figure}

In order to illustrate the discriminating power of the bin $[2,4.3]$ GeV$^2$ of $P_2$, we show on the left-hand side of Fig.~\ref{fig:P2max} the curve of $P_2$ (central value) in the neighbourhood of its maximum together with the integrated result for three different scenarios: the SM, a new physics scenario NP with $C_9^{\rm NP}=-1.5$, and a new physics scenario RHC with the right-handed currents $C_7^\prime=0.06$, $C_9^\prime=1$, $C_{10}^\prime=0.4$. In the scenario NP,  the maximum of $P_2$ is not lowered but its position is shifted to a higher $q^2$-value leading to a better agreement of the integrated result with the measured value. In the scenario RHC, on the other hand, the height of the maximum is lowered resulting in a stronger deviation of the integrated result from the measured value compared to the SM case. The scenario RHC has been chosen in such a way that the central values for all low-$q^2$ bins of $P_1$ fall within the experimental 1$\sigma$ regions, as demonstrated in the plot on the right-hand side of Fig.~\ref{fig:P2max}. It thus constitutes an illustrative example of a setup with new right-handed currents to which the maximum of $P_2$ exhibits a stronger sensitivity than the observable $P_1$.

\boldmath
\subsection{Relation between $P_4^\prime$ and $P_5^\prime$ at the position of maximum and at the zero of $P_2$}\unboldmath
\label{subsec:RelP45}
Eq.~(\ref{eq:PRel}) is quadratic in the parameters $P_4^\prime$,$P_5^\prime$,$P_6^\prime$,$P_8^\prime$. The requirement of real solutions for these observables constrains the allowed ranges of possible values. For example, demanding a real solution for $P_4^\prime$ from Eq.~(\ref{eq:PRel}) implies
\begin{equation}
   0\le\Delta(P_4^\prime)=-4x(P_5^\prime)^2-4x(P_8^\prime)^2-4\left[(k_T+P_1)P_6^\prime-2P_2P_8^\prime-2P_3P_5^\prime\right]^2+4xk_L(k_T-P_1), \label{eq:Delta4}
\end{equation}
with $x$ defined in eq.~(\ref{eq:obsx}) and fulfilling $x\ge 0$. Hence, the first three terms in eq.~(\ref{eq:Delta4}) are negative definite and each of them has thus to be smaller in absolute value than the positive fourth term. From this observation we can directly read off constraints on $|P_5^\prime|$ and $|P_8^\prime|$, while constraints on $|P_4^\prime|$ and $|P_6^\prime|$ can for example be obtained by considering $\Delta(P_5^\prime)$. The total set of constraints is given by
\begin{equation}
  |P_4^\prime|\le\sqrt{k_L(k_T-P_1)},\hspace{0.7cm}
  |P_5^\prime|\le\frac{1}{\beta}\sqrt{k_L(k_T+P_1)},\hspace{0.7cm}
  |P_6^\prime|\le\frac{1}{\beta}\sqrt{k_L(k_T-P_1)},\hspace{0.7cm}
  |P_8^\prime|\le\sqrt{k_L(k_T+P_1)}.
\end{equation}
As before, these bounds (with the reinstalled $\beta$-dependence for $P_5^\prime$ and $P_6^\prime$) are valid up to quadratic terms in CP-violating coefficients, while exact versions can be obtained via the replacement rules (\ref{eq:ExactRep1}) and (\ref{eq:ExactRep2}). The constraints are obtained for $x>0$ and thus are valid for any $q^2$ except for the single point where $x$ reaches its minimum value $x=0$. Continuity of the $P_i^\prime$ then implies the bounds to be valid also for $x=0$.

In the limit $x\to 0$ the third term in eq.~(\ref{eq:Delta4}) has to vanish in order to render $P_4^\prime$ real. Proceeding in the same way for the other $\Delta(P_{5,6,8}^\prime)$ we obtain four relations at $q^2=q_1^2$ with $x(q_1^2)=0$:
\begin{eqnarray}
  \left[(k_T+P_1)P_6^\prime-2P_2P_8^\prime-2P_3P_5^\prime\right]_{q_1^2}&=&0,\nn\\
  \left[(k_T-P_1)P_8^\prime-2P_2P_6^\prime+2P_3P_4^\prime\right]_{q_1^2}&=&0,\nn\\
  \left[(k_T+P_1)P_4^\prime-2P_2P_5^\prime+2P_3P_8^\prime\right]_{q_1^2}&=&0,\nn\\
  \left[(k_T-P_1)P_5^\prime-2P_2P_4^\prime-2P_3P_6^\prime\right]_{q_1^2}&=&0.
\end{eqnarray}
Neglecting $P_3P_{6,8}^\prime\ll P_2P_{4,5}^\prime$ and including the $\beta$-factor for $P_5^\prime$, the last two equations reduce to
\begin{equation}\label{eq:P45x}
   P_4^\prime(q_1^2)=\left[\beta P_5^\prime\sqrt{\frac{k_T-P_1}{k_T+P_1}}\right]_{q_1^2}.
\end{equation}
This relation is valid at the zero $q_1^2$ of $x$ where $P_2=\sqrt{k_T^2-P_1^2}/2\beta$. For $P_1\ll 1$, which is an excellent approximation in the absence of new right-handed currents, $q_1^2$ coincides with the position of the maximum $P_2^{\rm max}\approx k_T/(2\beta)$ of $P_2$, and Eq.~(\ref{eq:P45x}) becomes 
\begin{equation}\label{eq:P45max}
   P_4^\prime(q_1^2)\,=\,\beta(q_1^2) P_5^\prime(q_1^2).
\end{equation}
While Eq.~(\ref{eq:P45x}) is model-independent, Eq.~(\ref{eq:P45max}) only applies if there are no new right-handed currents. Its experimental validation therefore provides a test on the size of right-handed currents.

An analogous relation between $P_4^\prime$ and $P_5^\prime$ at the position $q^2=q_0^2$ of the zero of $P_2$ was derived and discussed in Ref.~\cite{nicola}. We reproduce it here for completeness. Dropping quadratic terms in $P_3$, $P_{6,8}$ and in the $P_i^{\rm CP}$ it reads
\begin{equation} \label{eq:P45zero} [{P_4}^{\prime 2} + \beta^2{P_5^{\prime 2}}]_{q_0^{2}}= 1 - \eta(q_0^2),\end{equation}
where $\eta(q_0^2)=[{ P_1}^2 +{ P_1} ({ P_4}^{\prime 2}-\beta^2 { P_5}^{\prime 2})]_{q_0^2}$ is completely negligible (of order $\eta(q_0^{2})\sim 10^{-3}$) in the absence of new right-handed currents. Let us assume that, as data seem to suggest, the zero $q_0^2$ of $P_2$ would be larger than predicted by the SM. In this case, Eq.~(\ref{eq:P45zero}) forces the absolute value of $P_5^\prime(q_0^2)$ to be smaller than in the SM, in agreement with the anomaly.

\begin{figure}
\includegraphics[width=8cm,height=5cm]{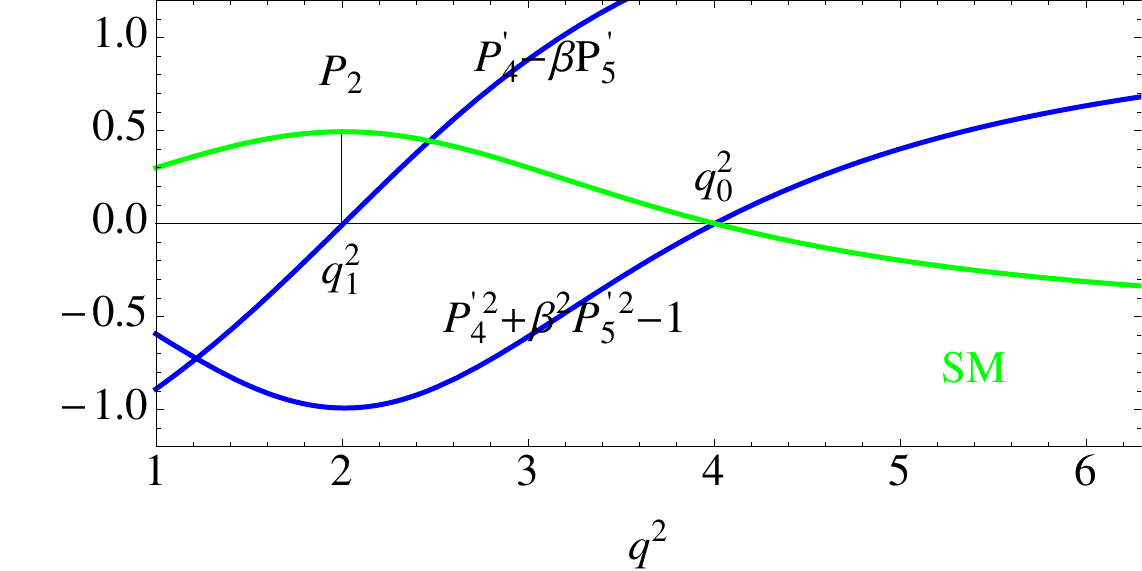}\hspace{0.5cm}
\includegraphics[width=8cm,height=5cm]{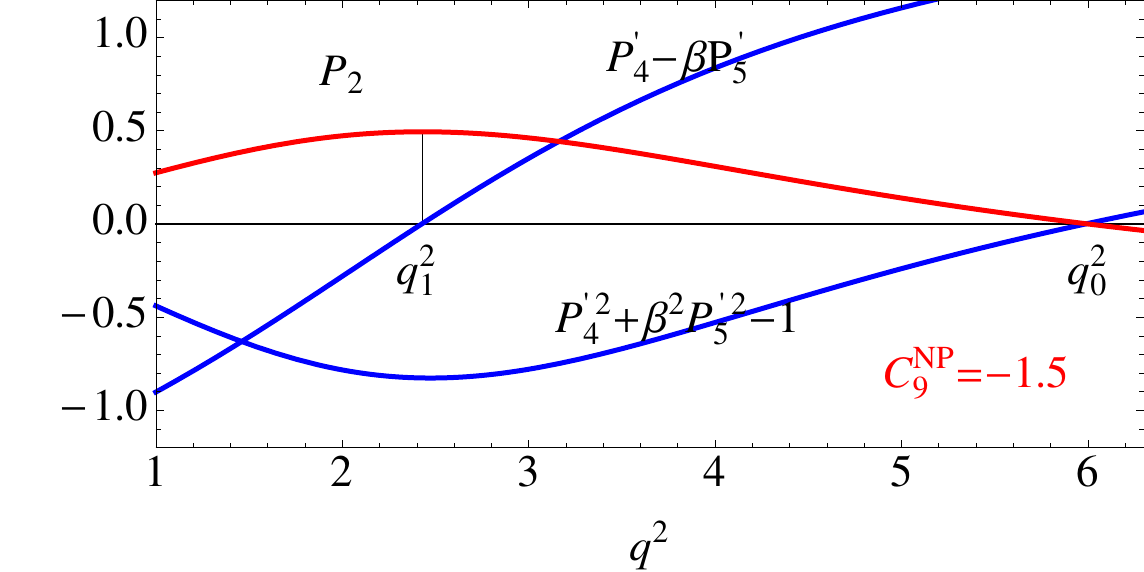}
\caption{\label{fig:relP45}Illustration of the relations (\ref{eq:P45max}) and (\ref{eq:P45zero}) between the observables
$P_4^\prime$ and $P_5^\prime$ (central values) at the position of the maximum and the zero of $P_2$. Left: SM. Right:  scenario NP with $C_9^{\rm NP}=-1.5$.}
\end{figure}

In Fig.~\ref{fig:relP45} we show central values of the theory predictions for the two functions $P_4^\prime-\beta P_5^\prime$ and $(P_4^\prime)^2+\beta^2(P_5^\prime)^2-1$ for the SM and the new-physics scenario NP with $C_9^{\rm NP}=-1.5$. The zeros of the corresponding curves at $q^2=q_1^2$ and $q^2=q_0^2$, respectively, demonstrate that the relations (\ref{eq:P45max}) and (\ref{eq:P45zero}) are indeed fulfilled to excellent precision. 

\boldmath
\subsection{Constraints on the $A_S^{(i)}$ and relations at the position of the maximum and the zero of $P_2$}\unboldmath
\label{subsec:As}
Eq.~(\ref{eq:AsRel1}) is quadratic in the parameters $A_S^4$,$A_S^5$,$A_S^7$,$A_S^8$. The requirement of real solutions for these observables constrains the allowed ranges of possible values. Following the procedure described in Sec.~\ref{subsec:RelP45} for the $P_i^\prime$, we find in a completely analogous manner the bounds
\begin{eqnarray}\label{eq:AsiBounds}
  &&|A_S^4|\le\frac{1}{2}\sqrt{3k_SF_TF_S(1-F_S)(k_T-P_1)},\hspace{2cm}
  |A_S^5|\le\sqrt{3k_SF_TF_S(1-F_S)(k_T+P_1)},\nn\\
  &&|A_S^7|\le\sqrt{3k_SF_TF_S(1-F_S)(k_T-P_1)},\hspace{2.2cm}
  |A_S^8|\le\frac{1}{2}\sqrt{3k_SF_TF_S(1-F_S)(k_T+P_1)}.
\end{eqnarray}
Combining further the Eqs.~(\ref{eq:PRel})-(\ref{eq:AsRel2}), one obtains a similar bound on $A_S$ (see Appendix~B for a detailed derivation):
\begin{equation}\label{eq:AsBound}
   |A_S|\le 2\sqrt{3k_Lk_SF_S(1-F_S)(1-F_T)}
\end{equation}
The constraints (\ref{eq:AsiBounds}) and (\ref{eq:AsBound}) are identical to the ones given in eq.~(51) of Ref.~\cite{optimized} up to the Breit-Wigner factor $F=Z/\sqrt{XY}$ present in the latter. The results of Ref.~\cite{optimized} were derived using a different method based on the Cauchy-Schwartz inequality. The factor $F$ is a consequence of the implicit assumption of a narrow S-wave resonance made in Ref.~\cite{optimized}, and it has to be replaced by its upper limit $F_{\rm max}=1$ in the general case. This subtlety has little impact on the numerical results given in Ref.~\cite{optimized} as the phenomenological analysis there was performed taking $F=0.9\approx 1$. We further note that once again the stated results in Eqs.~(\ref{eq:AsiBounds}) and (\ref{eq:AsBound}) are valid up to quadratic terms in CP-violating coefficients, with exact versions being obtained via the replacements (\ref{eq:ExactRep1}) and (\ref{eq:ExactRep2}).
\begin{figure}
\includegraphics[width=4.0cm,height=4.0cm]{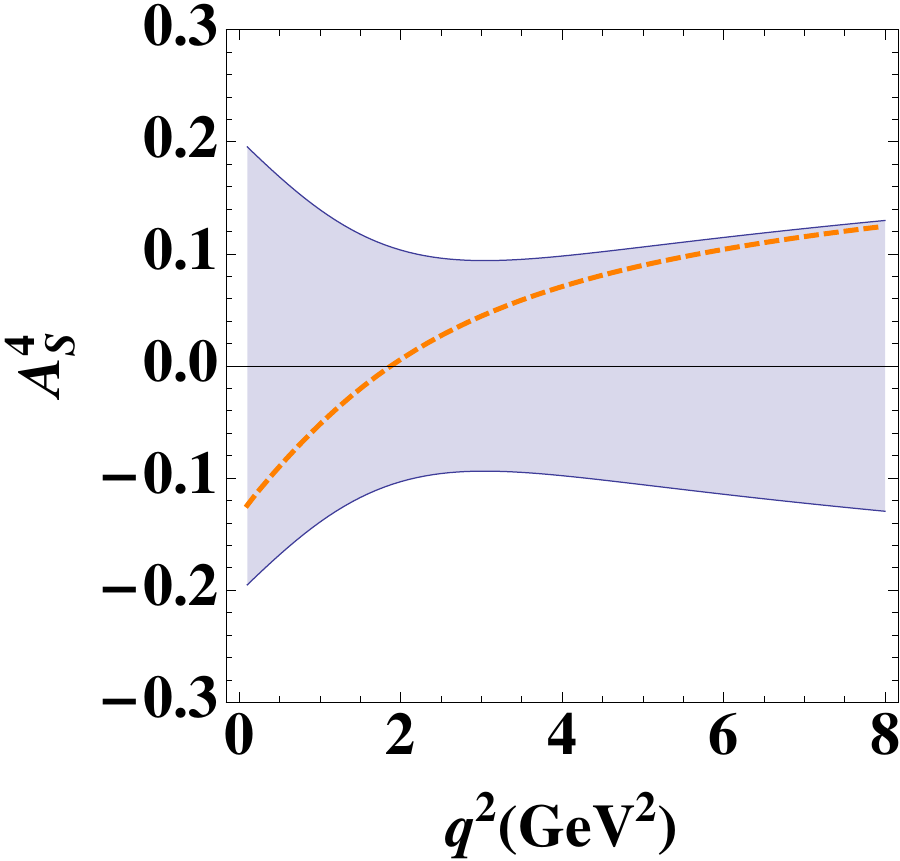}\hspace{0.3cm}
\includegraphics[width=4.0cm,height=4.0cm]{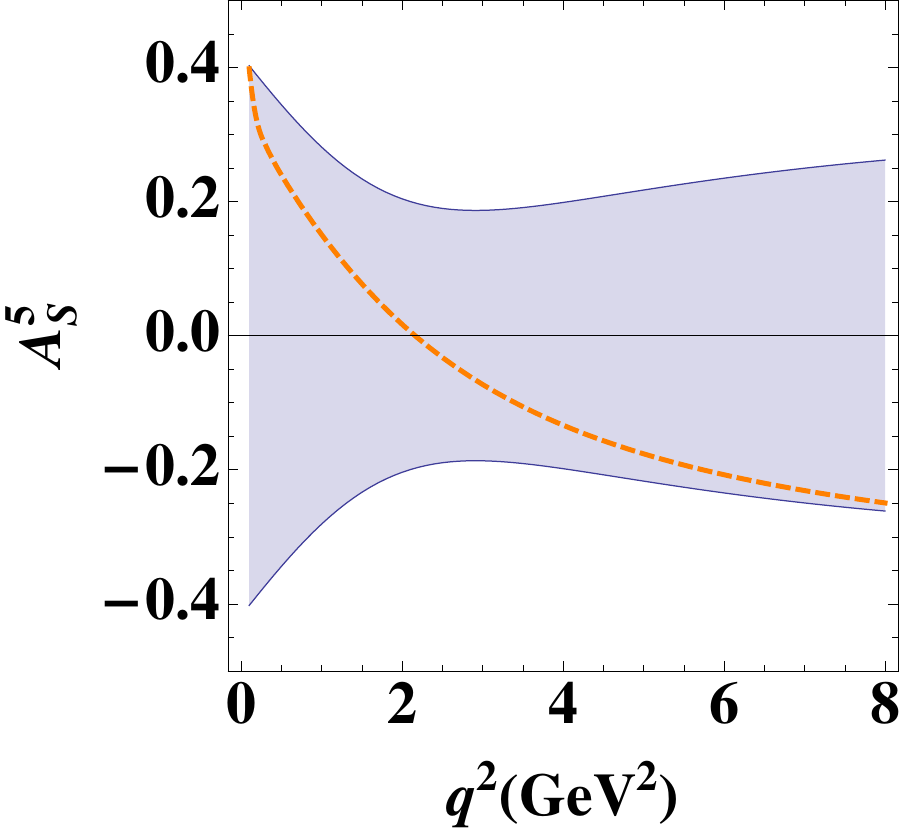}\hspace{0.3cm}
\includegraphics[width=4.0cm,height=4.0cm]{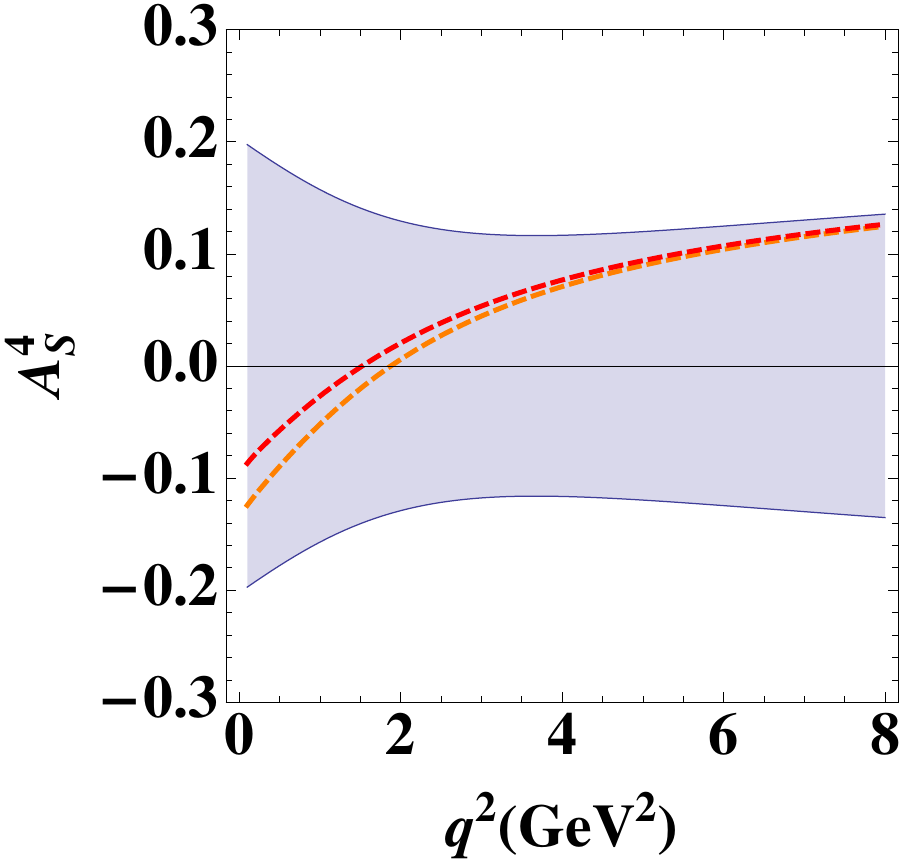}\hspace{0.3cm}
\includegraphics[width=4.0cm,height=4.0cm]{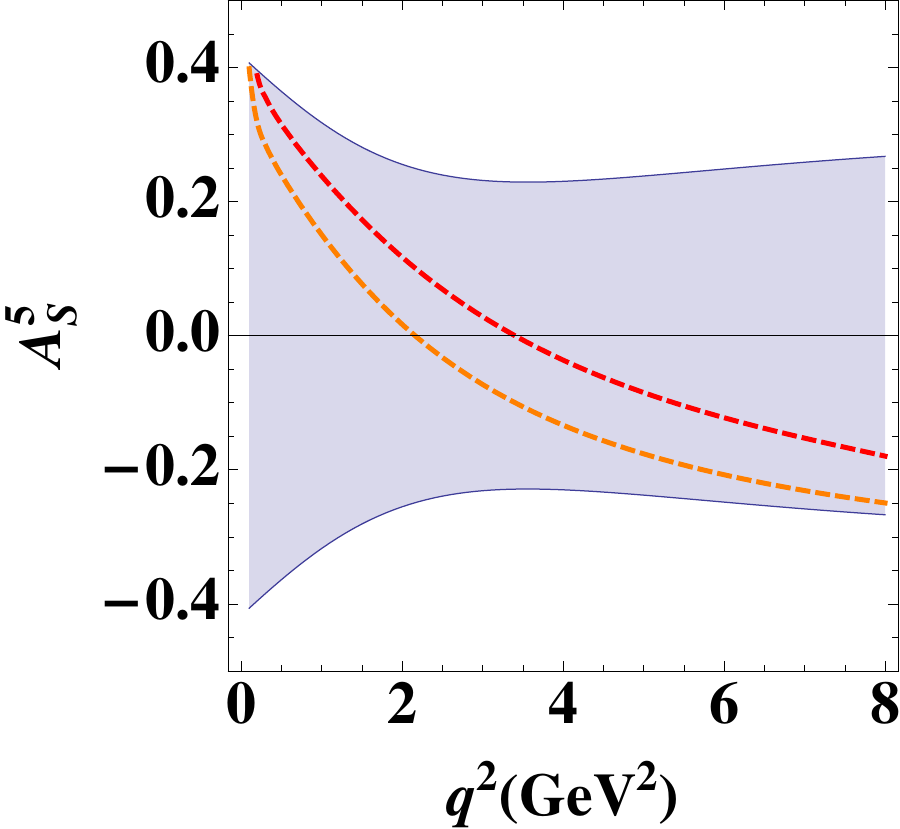}
\caption{\label{illus} Illustration of the constraints on $A_S^{4,5}$ obtained from relation (\ref{eq:AsRel1}) in
the SM (left two plots) and in the presence of $C_9^{\rm NP}=-1.5$ (right two plots).
Blue bands correspond to the uncorrelated bounds from eq.~(\ref{eq:AsiBounds}). Dashed lines illustrate
the correlation between $A_S^4$ and $A_S^5$ obtained from relation (\ref{eq:AsRel1}) for the scenario
described in the text (orange for the SM, red for $C_9^{\rm NP}=-1.5$). 
}
\end{figure}


Proceeding in an analogous way as in the P-wave case in Sec.~\ref{subsec:RelP45}, we find also for the S-wave parameters relations at the position $q^2=q_1^2$ of the zero of the observable $x$. The corresponding equations read
\begin{eqnarray}
  \left[(k_T+P_1)A_S^7-4P_2A_S^8+2P_3A_S^5\right]_{q_1^2}&=&0,\nn\\
  \left[(k_T-P_1)A_S^8-P_2A_S^7-2P_3A_S^4\right]_{q_1^2}&=&0,\nn\\
  \left[(k_T+P_1)A_S^4-P_2A_S^5-2P_3A_S^8\right]_{q_1^2}&=&0,\nn\\
  \left[(k_T-P_1)A_S^5-4P_2A_S^4+2P_3A_S^7\right]_{q_1^2}&=&0,
\end{eqnarray}
and simplify to
\begin{equation}\label{eq:AsRelx}
   2A_S^4(q_1^2)=\left[A_S^5\sqrt{\frac{k_T-P_1}{k_T+P_1}}\right]_{q_1^2}\hspace{1.5cm}\text{and}\hspace{1.5cm}
   A_S^7(q_1^2)=\left[2A_S^8\sqrt{\frac{k_T-P_1}{k_T+P_1}}\right]_{q_1^2}
\end{equation}
under the assumption of $P_3A_S^i\ll P_2A_S^j$. For $P_1\ll 1$, one obtains at the position $q_1^2$ of the maximum of $P_2$: 
\begin{equation}\label{eq:AsRelMax}
   2A_S^4(q_1^2)=A_S^5(q_1^2)\hspace{1.5cm}\text{and}\hspace{1.5cm}
   A_S^7(q_1^2)=2A_S^8(q_1^2).
\end{equation}

The symmetry relation (\ref{eq:AsRel1}), together with the implicitly contained relations  
(\ref{eq:AsRelx}),(\ref{eq:AsRelMax}) at the zero $q_1^2$ of $x$, imposes correlations among the $A_S^i$ implying constraints that go beyond the individual bounds given in Eqs.~(\ref{eq:AsiBounds}),(\ref{eq:AsBound}). To illustrate this, we assume that a measurement gives $A_S^{7,8}\ll A_S^{4,5}$. In this case, the symmetry relation (\ref{eq:AsRel1}) implies a direct correlation between $A_S^4$ and $A_S^5$. If for example $A_S^5$ is measured to be $A_S^5=\alpha P_5^\prime$ where $\alpha=\sqrt{3 F_T F_S (1-F_S)k_S/k_L}$, $A_S^4$ is completely fixed to $A_S^4=\frac{\alpha}{2} P_4^\prime$. This scenario is illustrated in
Fig.~\ref{illus} for constant $F_S \simeq 6\%$. The orange dashed curves in the plots on the left are obtained for SM values of
$P_{4,5}^ \prime$, while the red dashed curves in the plots on the right correspond to the presence of $C_9^{\rm NP}=-1.5$ (in addition the SM curve is shown also in the plots on the right to visualize the shift between the two curves). If one of the curves is measured for $A_S^5$, the corresponding curve for $A_S^4$ is predicted by the symmetry relation, and vice versa. Note that 
also the blue bands corresponding to the uncorrelated bounds from eq.~(\ref{eq:AsiBounds}) are slightly different in the SM and 
in the NP case.

As in the previous section for the P-wave observables, we give also for the S-wave observables simplified versions of the symmetry relations at the zero $q^2=q_0^2$ of $P_2$. Neglecting the small ${ P}_3, { P}_{6,8}^\prime $ terms, Eqs.~(\ref{eq:AsRel1}) and (\ref{eq:AsRel2}) simplify to 
\begin{equation}
[(4 A_S^{4 \, 2} + A_S^{7 \, 2}) (1 +P_1) + (A_S^{5 \, 2}+ 4 A_S^{8 \, 2}) (1-P_1)]_{q_0^2} = 3 [(1- F_S) F_S F_T (1- P_1^2)]_{q_0^2},
\end{equation}
\begin{equation}
A_S(q_0^2)= \left[\frac{ 2 \sqrt{1-F_T} (2 A_S^4 (1+ P_1) P_4^\prime + A_S^5 (1-P_1) P_5^\prime)}{\sqrt{F_T} (1-P_1^2)}\right]_{q_0^2}.
\end{equation}

\section{Conclusions}
\label{sec:Conclu}
In this article we have exploited the spin symmetry of the angular distribution of the decay $B\to K^*\mu^+\mu^-$, both in the P-wave as well as in the S-wave sector. We have shown that the symmetry reduces the number of independent S-wave observables from six to four, implying two non-trivial relations among the observables $F_S$, $A_S$, $A_S^4$, $A_S^5$, $A_S^7$ and $A_S^8$ which we derived explicitly. The relations allowed us to obtain individual bounds on the $A_S^{(i)}$ which agree with the ones determined in Ref.~\cite{optimized} via the Cauchy-Schwartz inequality. However, the constraining power of the symmetry relations goes beyond these individual bounds as they correlate the S-wave observables among each other. The implementation of these correlations into the experimental data analysis is expected to reduce the background from S-wave pollution. As an example, we have shown how for $A_S^{7,8}\ll A_S^{4,5}$ the correlations fix $A_S^4$ from a measurement of $A_S^5$ and $F_S$ (or $A_S^5$ from a measurement of $A_S^4$ and $F_S$) in the whole range of the of the squared dilepton invariant mass $q^2$.
 We further showed that $A_S^4/A_S^5$ and $A_S^7/A_S^8$ are completely fixed at a point $q^2=q_1^2$ where $q_1^2$ coincides with the position of the maximum of the P-wave observable $P_2$ in the absence of new right-handed currents.

We also pointed out the strong potential of the maximum of $P_2$ for probing NP beyong the SM, in particular the presence of new right-handed currents, in a region far away from charm resonances. We have shown that a shift of the position of the maximum of $P_2$ compared to its SM expectation, with the height of the maximum $P_2^{\rm max}$ kept at the SM value $1/(2\beta)$, would be a signal of a NP contribution to the SM-like Wilson coefficients $C_7$, $C_9$, $C_{10}$. A maximum value $P_2^{\rm max}<1/(2\beta)$, on the other hand, would detect the presence of new right-handed currents and thus complement information from the (currently not very precisely measured) observable $P_1$. We have further proven and illustrated that for $C_7^\prime=C_9^\prime=C_{10}^\prime=0$ the angular observables $P_4^\prime$ and $P_5^\prime$ fulfill $P_4^\prime=\beta P_5^\prime$ at the position of the maximum of $P_2$, so that any deviation from this relation would equally signal the presence of new right-handed currents.

\bigskip
{\it Acknowledgments.} We would like to thank K.~A.~Petridis and N.~Serra for many useful discussions. L.H. has been supported by FPA2011-25948 and the grant 2014 SGR 1450, and in part by the Centro de Excelencia Severo Ochoa SEV-2012-0234. J.M. acknowledges financial support from the Explora project FPA2014-61478-EXP.

\section*{Appendix A: Gauge conditions for the amplitudes}
\label{sec:AppA}

All the angular observables studied by the LHCb experiment are invariant under a $U(2)$ rotation
of the vectors $n_i$ defined in Eq.~(\ref{eq:nvecs}). As a consequence, the amplitudes $A_i^{L,R}$ cannot be determined unambigously from experiment. In order to arrive at a one-to-one correspondence between the experimental observables and the theoretical amplitudes, one has to fix a convention which picks for every class of $U(2)$-related amplitudes a certain representant (similar to "fixing the gauge"). One convenient choice that has been proposed and is used by the Imperial group of the LHCb experiment~\cite{Kostas} consists in requiring
$${\rm Re} A_0^R=0, \quad {\rm Im} A_0^R=0, \quad {\rm Im} A_0^L=0, \quad {\rm Im} A_\perp^R=0.$$
This choice is not unique, several combinations are possible (see Ref.~\cite{ulrik1} for a different choice). Starting from an arbitrary amplitude, one arrives at the above configuration by means of the $U(2)$ transformation
\eq{
  n_i \to
  \left[
    \begin{array}{ll}
      e^{i\phi_L} & 0 \\
      0 & e^{-i \phi_R}
    \end{array}
  \right]
  \left[
    \begin{array}{rr}
      \cos \theta & -\sin \theta \\
      \sin \theta &  \cos \theta
    \end{array}
  \right]
  \left[
    \begin{array}{rr}
      \cosh i \tilde{\theta} &  -\sinh i \tilde{\theta} \\
      - \sinh i \tilde{\theta} & \cosh i \tilde{\theta}
    \end{array}
  \right]
  n_i \,. \nonumber
 \label{symmassless}}
with
$${\tan 2{\tilde{\theta}}}= 2 \frac{{\rm Im} A_0^R {\rm Re} A_0^L + (L \leftrightarrow R)}{|A_0^R|^2-|A_0^L|^2}\,,$$
$${\tan \theta}=\frac{{\rm Re} A_0^R + {\rm Im} A_0^L {\rm tan}{\tilde\theta}}{-{\rm Re} A_0^L + {\rm Im} A_0^R {\rm tan}{\tilde\theta}}\,,$$
$${\tan \phi_L}=\frac{{\rm Im} A_0^L + {\rm Im} A_0^R \tan\theta - ({\rm Re} A_0^R - {\rm Re} A_0^L \tan\theta)\tan\tilde\theta}{-{\rm Re} A_0^L + {\rm Re} A_0^R \tan\theta + ({\rm Im} A_0^R  + {\rm Im} A_0^L \tan\theta)\tan\tilde\theta}\,,$$
$${\tan \phi_R}=\frac{{\rm Im} A_\perp^R + {\rm Im} A_\perp^L \tan\theta - ({\rm Re} A_\perp^L - {\rm Re} A_\perp^R \tan\theta)\tan\tilde\theta}{-{\rm Re} A_\perp^R + {\rm Re} A_\perp^L \tan\theta + ({\rm Im} A_\perp^L  + {\rm Im} A_\perp^R \tan\theta)\tan\tilde\theta}\,.$$

\boldmath
\section*{Appendix B: Derivation of the bound on $A_S$}
\label{sec:AppB}\unboldmath
In this appendix we present an explicit derivation of the constraint on the S-wave observable $A_S$ given in Eq.~(\ref{eq:AsBound}). Combining the relations (\ref{eq:PRel})-(\ref{eq:AsRel2}) as 
$a^2$(\ref{eq:PRel})$+3b^2$(\ref{eq:AsRel1})$+ab$(\ref{eq:AsRel2}) with arbitrary real coefficients $a,b$, one obtains an equation for linear combinations $aP_i^\prime\pm(2)bA_S^i$ of P- and S-wave observables which has the same structure as the individual $P_i^\prime$- and $A_S^i$-relations (\ref{eq:PRel}) and (\ref{eq:AsRel1}):
\begin{eqnarray}
   Y(a,b)\left[k_T^2-P_1^2-4P_2^2-4P_3^2\right]&=&-4P_2\left[(aP_4^\prime+2bA_S^4)(aP_5^\prime+bA_S^5)+(aP_6^\prime-bA_S^7)(aP_8^\prime-2bA_S^8)\right]\nonumber\\
       &&-4P_3\left[(aP_5^\prime+bA_S^5)(aP_6^\prime-bA_S^7)-
          (aP_4^\prime+2bA_S^4)(P_8^\prime-2bA_S^8)\right]\nonumber\\
       &&+(k_T+P_1)\left[(aP_4^\prime+2bA_S^4)^2+(aP_6^\prime-bA_S^7)^2\right]\nonumber\\
       &&+(k_T-P_1)\left[(aP_5^\prime+bA_S^5)^2+(aP_8^\prime-2A_S^8)^2\right],\label{eq:PAsRel}
\end{eqnarray}
with
\begin{eqnarray}\label{eq:Yab}
  Y(a,b)&=&a^2k_L+3b^2k_SF_TF_S(1-F_S)+abA_S\sqrt{\frac{F_S}{1-F_S}}\nonumber\\
        &=&k_L\left[a+\frac{b}{2k_L}A_S\sqrt{\frac{F_T}{1-F_T}}\right]^2
           \,+\,\frac{b^2}{4k_L}\frac{F_T}{1-F_T}\left[12k_Lk_SF_S(1-F_S)(1-F_T)-A_S^2\right].
\end{eqnarray}
Requiring $\Delta(aP_4^\prime+2bA_S^4)\ge 0$ in analogy to Eq.~(\ref{eq:Delta4}) in order to ensure that the observable $aP_4^\prime+2bA_S^4$ is real, one finds that $Y(a,b)\ge 0$. This condition has to be fulfilled for any possible linear combination, i.e. for any value of $a,b$, which according to Eq.~(\ref{eq:Yab}) enforces $A_S$ to respect the constraint (\ref{eq:AsBound}):
\begin{equation}
   |A_S|\le 2\sqrt{3k_Lk_SF_S(1-F_S)(1-F_T)}.
\end{equation}


\end{document}